\documentstyle[amssymb,twocolumn,prb,aps]{revtex}

\input{psfig.sty}

\begin{document}
\draft

\twocolumn[
\hsize\textwidth\columnwidth\hsize\csname @twocolumnfalse\endcsname
\title{Interplay between spin-relaxation and Andreev reflection in ferromagnetic wires 
with superconducting contacts}
\author{Vladimir I. Fal'ko $^{\ast}$, A.F. Volkov $^{\dagger \ast}$, and 
C. Lambert $^{\ast}$}
\address{$^{\ast }$ School of Physics and Chemistry, Lancaster University, LA1 4YB, UK\\
$^{\dagger }$ Institute for Radio-Engineering and Electronics, Moscow,  Russia}
\date{today}
\maketitle

\begin{abstract}
We analyze the change in the resistance of a junction between a
diffusive ferromagnetic (F) wire and normal metal electrode, due 
to the onset of superconductivity (S) in the latter and a double Andreev
scattering process leading to a complete internal reflection of
a large fraction of the spin-polarized electrons back into the ferromagnet. 
The superconducting transition results in an additional contact 
resistance arising from the necessity to match 
spin-polarized current in F-wire to spin-less current in S-reservoir, 
which is comparable to the resistance of a piece of a 
F-wire with the length equal to the spin-relaxation length.
\end{abstract}


\narrowtext]

Recent advances in microprocessing technologies of metals have allowed one
to obtain high quality nanostrucutres formed from combinations of
superconductoring (S) and ferromagnetic (F) materials \cite
{Petrashov1,Giordano,Pannetier,Vasko,Upadhyay}. In such structures one may
expect to see manifestations of Andreev reflection \cite{Andreev} at the FS
interface whereby electrons with\ excitation energies $\varepsilon \lesssim
\Delta $ from a normal (N) metal convert into a hole with the opposite spin,
and the normal current in the N-part of a circuit transforms into
supercurrent inside the superconductor. Both theoretical and experimental
studies of subgap transport of NS structures have revealed that Andreev
reflection may substantially change the circuit resistance - by the values
comparable to the resistance of a mesoscopic length segment of the normal
metal wire \cite
{Petrashov2,Vegvar,Esteve,Wees,Pannetier2,Review1,Review2,Review3,Nazarov,Zaitsev}%
. The difference between the NN and NS resistances, $R_{{\rm N}}$ and $R_{%
{\rm S}}$, is determined by the extent of the proximity effect over the
non-superconducting material. In the presence of a large exchange field $%
\varepsilon _{ex}\gg \Delta $, electron-hole correlations are suppressed in
ferromagnets on a microscopic length scale, so that Andreev processes at the
FS interface do not generate subgap conductance effects related to
condensate penetration into the non-superconducting part of the circuit.

However, there exists another mechanism of large conductance variations
below the critical temperature $T_{c}$ in FS structures, which has a
classical nature and arises from the necessity to match the spin-polarized
electron current at one end of the ferromagnetic lead to the spinless
current inside the superconductor, whereas above $T_{c}$ the boundary
conditions allow the spin-polarized current to flow through the entire
circuit. The degree of spin polarization, $\varsigma =\left( j_{\uparrow
}-j_{\downarrow }\right) /\left( j_{\uparrow }+j_{\downarrow }\right) $
carried by the electric current in a free standing single magnetic domain
F-wire, 
\begin{equation}
\varsigma =\left( D_{+}\nu _{+}-D_{-}\nu _{-}\right) /\left( D_{+}\nu
_{+}+D_{-}\nu _{-}\right) ,  \label{pd}
\end{equation}
may be substantial if Fermi-surfaces of spin-up and spin-down electrons are
very different (e.g., $D_{+}\nu _{+}\ll D_{-}\nu _{-}$). Here, $D_{\alpha }=%
\frac{1}{3}v_{\alpha }l_{\alpha }$ and $\nu _{\alpha }$ are the diffusion
coefficients and densities of states in the ferromagnet, $\alpha =\pm $
describe spin 'up' and 'down', $v_{\alpha }$ are the Fermi velocities, $%
l_{\alpha }$ - the mean free paths. Note that we are interested in a
situation where the resistivity of ferromagnet dominates the circuit
resistance, and the resistance of the S-part of the system can be neglected
even when it is in the normal state. In the absence of spin-relaxation, the
only solution to this problem consists of assuming a slight non-equilibrium
(current-induced) spin-repolarization of the ferromagnetic wire, so that the
diffusion of locally accumulated non-equilibrium spin-density would
compensate the spin carried by the electric current. That is, everywhere
across the F-wire, the local values of chemical potentials of spin-up and
-down electrons split, which limits the conduction by that of the worst
conducting spin-state. By taking into account both the electron and
Andreev-reflected hole currents, this logical exercise may be upgraded to
yield an estimate of the resistance variation of a spin-conserving F-wire,\ $%
\left( R_{{\rm S}}-R_{{\rm N}}\right) /R_{{\rm N}}=(D_{+}\nu _{+}-D_{-}\nu
_{-})^{2}/4(D_{+}\nu _{+}D_{-}\nu _{-})\equiv \varsigma ^{2}/(1-\varsigma
^{2})$, which can be also deduced from the result obtained by de Jong and
Beenakker \cite{Beenakker} using the Landauer-Buttiker approach extended to
the hybrid NS structures \cite{Review1,Review3}.

In this paper, we study the evolution of the resistance of a ferromagnetic
wire under the superconducting transition in the bulk electrode with
emphasis on the case of a diffusive F-wire whose length is comparable to, or
much longer than the length $L_{s}$ of spin-relaxation processes. The
resistance of a macroscopically long wire (with cross-section $L_{\bot
}\times d$ and resistance per square $R_{\square }^{-1}=e^{2}(D_{-}\nu
_{-}+D_{+}\nu _{+})d$), can be split into a 'bulk' part and a contact
resistance $r_{{\rm c}}$ formed within mesoscopic region near the FN or FS
junction, such that $R=\left( L/L_{\bot }\right) R_{\square }+r_{{\rm c}}$.
As we find below, the contact resistance $r_{{\rm c}}$ (which, in the normal
state, is determined by the relation between Fermi surfaces of carriers in F
and N) acquires below $T_{c}$ an additional contribution equal to the
resistance of a segment of the F-wire with the length$\ L_{s}$. Below, we
present the semiclassical analysis of this effect, which includes
calculation of the classical resistance variations near $T_{c}$ and down to
the zero-temperature limit, and an estimate of the weak localization
correction to it.

The resistance of a disordered F-wire can be found by solving diffusion
equations for the isotropic part of the electron distribution function, $%
n_{\alpha }(z,\varepsilon )=\int d\Omega _{{\bf p}}n_{\alpha }(z,{\bf p})$.
Due to the electron-hole symmetry, and in order to simplify the calculation
of the FS case, we shall compute the symmetrized function $N_{\alpha
}(\varepsilon ,z)=\frac{1}{2}\left[ n_{\alpha }(z,\varepsilon )+n_{\alpha
}(z,-\varepsilon )\right] $, where $\varepsilon $ is determined with respect
to the chemical potential in the S(N) electrode. In terms of $N_{\alpha
}(\varepsilon ,z)$, the current density is given by $j_{\alpha }=e^{2}\nu
_{\alpha }D_{\alpha }\int_{0}^{\infty }d\varepsilon \partial _{z}N_{\alpha
}(\varepsilon ,z)$, and $N_{\alpha }(\varepsilon ,z)$ obey the diffusion
equation 
\begin{equation}
D_{\alpha }\partial _{z}^{2}N_{\alpha }(z,\varepsilon )=w_{\uparrow
\downarrow }\nu _{-\alpha }\left[ N_{\alpha }(z,\varepsilon )-N_{-\alpha
}(z,\varepsilon )\right] ,  \label{diff1}
\end{equation}
\ which is more convenient to use in the equivalent form 
\begin{equation}
\partial _{z}^{2}\sum_{\alpha =\pm }D_{\alpha }\nu _{\alpha }N_{\alpha
}=0,\; \left[ \partial _{z}^{2}-L_{s}^{-2}\right] \left( N_{+}-N_{-}\right)
=0.  \nonumber
\end{equation}
The term on the right hand side of Eq. (\ref{diff1}) accounts for
spin-relaxation which may result from both spin-orbit scattering at defects
and the surface, and from the random weak precession of electron spins when
passing through adjacent non-collinearly magnetized ferromagnetic domains.
It can be used to define the effective spin-relaxation length, $L_{s}$ as $%
L_{s}^{-2}=w_{\uparrow \downarrow }\left[ \nu _{\uparrow }/D_{\downarrow
}+\nu _{\downarrow }/D_{\uparrow }\right] $. This pair of equations, which
ignore any energy relaxation, should be complemented by four boundary
conditions, two on each side of the ferromagnetic wire.

\begin{figure}[tbp]
\centerline{\psfig{figure=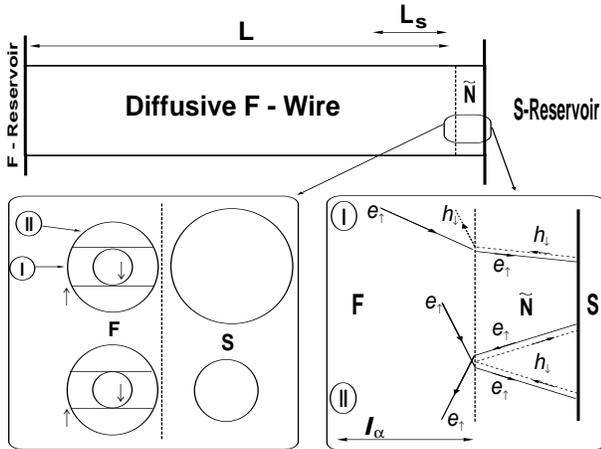,height=6cm,width=8cm}}
\caption{Pictural representation of the FS junction, a double Andreev
reflection processes in it, and of possible relations between Fermi surfaces of
spin-up and -down electrons in the F-wire (left) and in the N/S metal
(right) }
\label{fig1}
\end{figure}

To obtain boundary conditions for Eqs. (\ref{diff1}) and their solutions in
a specific geometry, we employ the model shown in Fig. 1, where the FS
junction is replaced by a sandwich of three layers: (i) a ferromagnetic (F)
wire of the length $L$ connected to the bulk F reservoir, (ii) a normal
metal layer ($\widetilde{N}$) which never undergoes a superconducting
transition by itself and has a negligible resistance, and (iii) a bulk
electrode S(N) which undergoes the superconducting transition. The insertion
of a normal metal layer $\widetilde{N}$ between the F and S(N) parts allows
us to formulate the boundary conditions at the FS interface using known
boundary conditions at the $\widetilde{N}$S interface \cite
{Review3,LarkinOvchinnikov}. For the sake of simplicity, we consider $%
\widetilde{N}$ as ballistic and the F$\widetilde{N}$ junction - as
semiclassically transparent, so that electrons either pass from one side to
the other, or are fully reflected, depending on whether this process is
allowed by the energy-momentum conservation near the Fermi surface. The
latter approximation avoids resonances through the 'surface states' \cite
{IgorFalko} due to multiple passage through the normal layer inserted
between S and F. As illustrated in Fig. 1, we approximate the spectrum of
electrons by parabolic bands - two for spin-down and spin-up electrons in F,
and one in the N-part, which we take into account by introducing the
parameters $\delta _{\alpha {\rm N}}^{2}=p_{{\rm FN}}^{2}/p_{{\rm F}\alpha
}^{2}$ and $\delta ^{2}=\left( p_{{\rm F}-}/p_{{\rm F}+}\right) ^{2}<1$. The 
$\widetilde{N}$N interface is assumed to be ideal, and the Fermi surfaces in 
$\widetilde{N}$ and N layers to be the same, so that $\widetilde{N}$S
Andreev reflection has unit probability. In such a model, the momentum of an
electron in the plane of the junction is conserved.

The boundary conditions on the left end are\ given by the equilibrium
distribution of electrons in the F-electrode, $N_{\alpha }(-L,\varepsilon )=%
\frac{1}{2}\left[ n_{T}(\varepsilon -eV)+n_{T}(-\varepsilon -eV)\right] $.
The boundary condition on the other end depends on the state of the
electrode, and in the superconducting state takes into account Andreev
reflection at the NS interface \cite{Andreev}. Since in our model of an
ideal F$\widetilde{N}$ interface, the parallel component of the electron
momentum is conserved, the effective reflection/transmission of electrons in
parts I and II of the ferromagnet Fermi surface sketched in Fig. 1 are
different. Although non-equilibrium quasi-particles from F pass inside $%
\widetilde{N}$ and generate holes by being Andreev reflected at the $%
\widetilde{N}$S interface, only those holes which are created by
quasi-electrons from part I of the Fermi surface in F may escape into the
F-wire. The spin-down holes which were generated by spin-up electrons from
part II of the Fermi surface cannot find states in F, so that they are fully
internally reflected into $\widetilde{N}$. Then, they undergo a second
Andreev reflection, convert into the spin-up electrons, and return back into
the ferromagnetic wire. This results in {\it complete internal reflection}
of spin-up electrons from part II of the Fermi surface inside the F-wire,
which nullifies the spin current through its FS edge.

The boundary condition near the F$\widetilde{N}$ junction can be found by
matching the isoenergetic electron fluxes determined in the diffusive region
found in the ballistic F-region using the reflection/transmission relation
between the distributions of incident and Andreev or normal reflected
electrons. The algebraic procedure used in this derivation is sketched in
footnote \cite{Footnote}. For quasi-particles with energies $0<\varepsilon
<\Delta $ this can be written in the form 
\begin{eqnarray}
D_{+}\nu _{+}\partial _{z}N_{+}-D_{-}\nu _{-}\partial _{z}N_{-} &=&0,
\label{LO} \\
N_{+}+N_{-}+\frac{2}{3}\varkappa \delta ^{2}l_{-}\partial _{z}N_{-}
&=&2N_{T}\left( \varepsilon \right) ,
\end{eqnarray}
where $\varkappa =(1-\delta ^{2})^{3/2}/\delta ^{2},\;\delta ^{2}=p_{{\rm F}%
-}^{2}/p_{{\rm F}+}^{2}<1$, and $N_{T}(\varepsilon )=\frac{1}{2}\left[
n_{T}(\varepsilon )+n_{T}(-\varepsilon )\right] =\frac{1}{2}$. A similar
result \cite{Footnote1} can be derived by considering the\ inserted $%
\widetilde{N}$-layer as diffusive, if we employ the known boundary
conditions from Refs. \cite{Review2,Review3,VolkovFalkounpub}. At $%
\varepsilon >\Delta $ the boundary conditions are the same as in the normal
state, $N_{\pm }=\frac{1}{2}$.

By solving them at low temperatures, $T\ll T_{c}$, we arrive at the contact
resistance of the FS boundary

\begin{equation}
r_{{\rm c}}^{{\rm S}}=R_{\square }\frac{L_{s}}{L_{\bot }}\frac{\varsigma ^{2}%
}{1-\varsigma ^{2}}+\frac{R_{\square }l_{+}}{3L_{\bot }}\frac{\varkappa }{%
1+\varsigma }.  \label{RcS}
\end{equation}

In the normal state of the right hand reservoir, the boundary conditions at
the end of F-wire depend on the relation between the Fermi momenta of
electrons in the ferromagnet and normal metal, 
\[
\left. N_{\alpha }(z,\varepsilon )+\frac{4\varkappa _{\alpha {\rm N}%
}D_{\alpha }}{v_{\alpha }}\partial _{z}N_{\alpha }(z,\varepsilon )\right|
_{z=0}=N_{T}(\varepsilon ), 
\]
where $\varkappa _{\alpha {\rm N}}=(1-\delta _{\alpha {\rm N}%
}^{2})^{3/2}/\delta _{\alpha {\rm N}}^{2},\;\delta _{\alpha {\rm N}}<1$, and 
$\varkappa _{\alpha {\rm N}}=0,\;\delta _{\alpha {\rm N}}\geq 1$, $\delta
_{\alpha {\rm N}}^{2}=p_{{\rm FN}}^{2}/p_{{\rm F}\alpha }^{2}$. These result
in the contact resistance term 
\begin{equation}
r_{{\rm c}}^{{\rm N}}=R_{\square }\frac{l_{+}\left( 1+\varsigma \right) }{%
L_{\bot }}\left\{ \left( 1-\varsigma \right) l_{+}/L_{s}+\frac{3}{2}%
\varkappa _{+{\rm N}}^{-1}\right\} ^{-1},  \label{RcN}
\end{equation}
which has sense only when it is larger than the resistance of the short
piece of the F-wire with the length of the order of $l_{+}$ and otherwise
should be neglected.

After comparing the latter result to $r_{{\rm c}}^{{\rm S}}$, we find that
the resistance of a long ferromagnetic wire attached to the S-electrode
exceeds the resistance of the same wire connected to the normal reservoir by
the resistance of a F-segment of length of order of $L_{s}$. One can extend
the result of Eq. (\ref{RcS}) to finite temperatures, which yields the
resistance variation below the superconducting transition 
\begin{equation}
R_{{\rm S}}(T)-R_{{\rm N}}\approx \frac{\varsigma ^{2}}{1-\varsigma ^{2}}%
\frac{L_{s}}{L_{\bot }}R_{\square }\tanh \left( \frac{\Delta (T)}{2T}\right)
.  \label{RSN}
\end{equation}
The increase of the resistance in Eq. (\ref{RSN}) originates from {\it the
matching of a spin-polarized current in the highly resistive ferromagnetic
wire to a spinless current inside the superconductor}, and represents the
main result of this paper. This robust classical effect is peculiar to
mono-domain wires, with the domain size $L_{{\rm D}}>L_{s}$. In a
multi-domain wire, with a finely coarse-grained collinear magnetic
structure, the transport properties of spin-up and down electrons do not
differ, $\varsigma \rightarrow 0$, the spin-current in the bulk of the
F-wire is equal to zero, and, therefore, the classical contact resistance
effect described by Eq. (\ref{RSN}) is absent.

When speaking about the opposite limit of a short wire with $L\ll L_{s}$, it
is more appropriate to discuss the conductance variation of the entire wire, 
$G_{{\rm N}}-G_{{\rm S}}$ rather than contact resistances, which can be
represented as 
\begin{equation}
\frac{G_{{\rm S}}(T)-G_{{\rm N}}}{G_{{\rm N}}}\approx \varsigma ^{2}\tanh
\left( \frac{\Delta (T)}{2T}\right) .
\end{equation}

In addition to the effect described by Eq. (\ref{RSN}) which originates from
a mesoscopic region of the F-wire, Eqs. (\ref{RcS}) and (\ref{RcN}) also
contain a small contribution to the variation $r_{{\rm c}}^{{\rm S}}-r_{{\rm %
c}}^{{\rm N}}$ from 2e charge transfer at the FS interface, due to the
Andreev process. The latter contribution to the classical resistance should
be taken into account only if its value greatly exceeds $R_{\square
}l_{+}/L_{\bot }$, and, for $\varsigma =0$, it has the form $\lim_{\varsigma
\rightarrow 0}\left( R_{{\rm S}}-R_{{\rm N}}\right) \approx \left[ \varkappa
-2\varkappa _{+{\rm N}}\right] R_{\square }l_{+}/3L_{\bot }$, similar to de
Jong and Beenakker's result for the FS ballistic point contact \cite
{Beenakker}. In contrast with the added contact resistance in Eq. (\ref{RSN}%
), this may have an arbitrary sign, depending on the ratios $\delta ^{2}\ $%
and $\delta _{+{\rm N}}^{2}$ between the areas of Fermi surfaces of
electrons in the normal metal and ferromagnet (for the system illustrated by
Fig. 1, $\varkappa >\varkappa _{+{\rm N}}$).

\begin{figure}[tbp]
\centerline{\psfig{figure=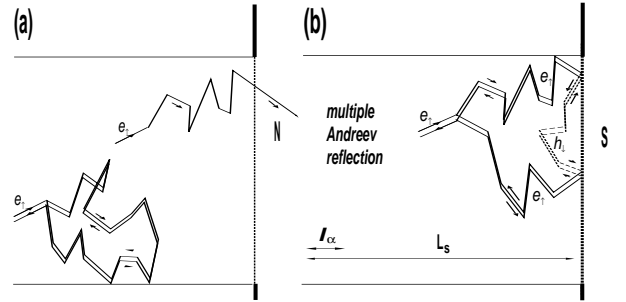,height=4cm,width=8cm}}
\caption{(a) Cooperon decay at the FN boundary due to the electron escape
into the normal reservoir. (b) Change of the boundary conditions for the
Cooperon due to a multiple Andreev reflection.}
\label{fig2}
\end{figure}

As wellas the above classical resistance effect, one may discuss the
variation of the weak localization correction to the contact resistance, in
particular, in wires with a finely coarse-grained collinear magnetic
structure, where Eq. (\ref{RSN}) gives zero effect. Since the weak
localization correction can be easily destroyed by a magnetic field, we have
to assume that the wire cross-section is small and the intrinsic magnetic
field in it is not large enough to suppress the enhanced back scattering
effect. To calculate the weak localization correction, $\Delta R_{{\rm S}%
}-\Delta R_{{\rm N}}$ to the variation of the contact resistance in the
latter situation, one should take into account the following features of the
problem: (a) The weak localization correction to the resistance, is
dominated by the triplet channel of the Cooperon, where the Cooperon spin
projection onto the overall magnetization direction is $m=\pm 1$, and is
restricted within the length scale $L_{s}$ (we assume that $L_{s}>L_{\bot }$%
). (b) The weak localization effect in the conductance is affected by the
change of the boundary conditions for the Cooperon due to the multiple
Andreev process, as illustrated in Fig. 2. In the normal state, the
electrons escape to the electrode, whereas in the S-state they undergo
several Andreev reflections, so that they may return to the same point
carrying the initial spin and contribute to interference. As a result, the
resistance of the entire circuit changes (between $T_{c}$ and a low
temperature) by an amount corresponding to the change of the conductance of
the last $L_{s}$-segment of the F-wire: 
\[
\Delta R_{{\rm S}}-\Delta R_{{\rm N}}\sim \frac{e^{2}}{h}\left( R_{\square }%
\frac{L_{s}}{L_{\bot }}\right) ^{2}. 
\]
The details of analogous estimations for a single-domain F-wire will be
reported elsewhere.

In conclusion, we have calculated the resistance variation of an FS
structure below the critical temperature and shown that the sign and
magnitude of this effect crucially depends on the domain structure of the
ferromagnetic part of the circuit. In a single domain F-wire with length
larger than spin-relaxation length $L_{s}$, this variation has classical
origin and is formed within the last $L_{s}$-segment of the wire (where the
spin-polarized current brought from the F-part relaxes into spin-less
current in a superconductor), and $R_{{\rm S}}(T)-R_{{\rm N}}$ increases
from zero at $T_{c}$ to a positive value at $T=0$. In a coarse-grained
multi-domain wire, $R_{{\rm S}}(T)-R_{{\rm N}}$ is determined by the
interplay between the mismatch of Fermi-surfaces and the weak localization
effect, and may have an arbitrary sign.

Authors thank V.Petrashov and R.Raimondi for discussions. This work was
funded by EPSRC and EC TMR Programme.


\begin{references}
\bibitem{Footnote1}  In particular, Eq. (\ref{LO}) is analogous to the
Larkin-Ovchinnikov boundary condition at the NS interface, $\partial _{z}%
\left[ n_{\pm }(\varepsilon )-n_{\mp }(-\varepsilon )\right] =0$ .
\end{references}
\end{document}